\documentclass[options]{JHEP3}
\title{Scale dependence of the power spectrum of the curvature perturbation determined using a numerical method in slow-roll inflation}
\author{Shiro Hirai\\ 
Department of Digital Games, Osaka Electro-Communication University\\
1130-70 Kiyotaki, Shijonawate, Osaka 575-0063, Japan\\ 
E-mail: \email{hirai@isc.osakac.ac.jp}}
\author{Tomoyuki Takami\\ 
Department of Digital Games, Osaka Electro-Communication University\\
1130-70 Kiyotaki, Shijonawate, Osaka 575-0063, Japan\\
E-mail: \email{takami@isc.osakac.ac.jp}}
\abstract{The Taylor expansion method has been used to investigate the scale dependence of the power spectrum of the curvature perturbation. In the present study, an alternative numerical method is used to clarify the $k$ dependence. Although there is thought to be no large difference between these two methods, some differences arise among various inflation models. For example, at $k$ = 1 Mpc, there is a 1.4 \% difference in the power spectrum, and with respect to the angular power spectrum, the difference of the value of $\chi^2$ nearly 10 occur in new inflation. However, in hybrid inflation, these differences do not occur. The time dependence of the inflationary and cosmological parameters is investigated, and differences among inflation models are clarified.}

\preprint{\today}
\dedicated{...}
\usepackage[dvips]{graphicx}

\begin{document}
\section{Introduction}
The curvature perturbation produced during the inflation epoch is an important quantity in cosmology. The scale dependence of the power spectrum of the curvature perturbation has been investigated in detail using the Taylor expansion around a pivot scale [1]. The scale dependence of the power spectrum can be written in terms of the slow-roll parameters and can be calculated in terms of higher orders of the slow-roll parameters [2]. In the present paper, a numerical method is proposed as an alternative method by which to investigate the $k$ dependence of the power spectrum. The proposed method is described in detail herein. In the case of a numerical method, the time dependence of the inflaton must be known in order to investigate the $k$ dependence of the power spectrum. In a previous study, we investigated the effect of the length of inflation [3]. However, in the present study, we assume that the length of inflation is very long, and so do not consider the contribution of the length of inflation. The time-dependent behavior of the inflaton during the last 100 $e$-folds in inflation is investigated using simple inflation models, because the physically interesting period of inflation is that in the last 60 $e$-folds. Although a correct inflationary potential based on the superstring or supergravity theory has yet to be established, the potential term in this latter period of inflation can be expressed in a simple form, such as the new inflation model, the chaotic inflation model, or the hybrid inflation model. Here, these three inflation models are considered in the present study. Next, using the derived time-dependent behavior of the inflaton, the time dependence of the slow-roll parameters, the Hubble parameter, and the spectral index can be calculated. The difference in time-dependent behavior of such parameters among the inflation models can then be investigated. Next, using the time dependence of the slow-roll parameters and the Hubble parameter, the differential equation of the gauge potential is solved by a numerical method. Finally, the $k$ dependence of the power spectrum of the curvature perturbation is derived using this solution. In order to investigate the difference between the familiar Taylor expansion and the newly proposed numerical method, the angular power spectrum and the value of $\chi^2$ are calculated.
\section{Cosmological and inflationary parameters}

The dependence of the cosmological and inflationary parameters on time is investigated using a slow-roll inflation model. Assuming a spatially flat universe and the energy density of the universe to be dominated by the inflaton field, the Einstein field equations can be written as 
\begin{equation}
H^2 = \frac{8\pi}{3m^2}(V(\phi)+\frac{1}{2}\dot{\phi}^2),
\end{equation}
and
\begin{equation}
\ddot{\phi}+3H\dot{\phi}+V^\prime(\phi)=0,
\end{equation}
where $\phi$ is the inflaton field, $V(\phi)$ is the inflaton potential, $m$ is the Planck mass, and $H=\dot{a}/a$ with $a$ being a scale factor. The variable $N$ which is defined by $dN = Hdt$, is introduced to represent time with respect to the number of $e$-folds of inflation. Equation (2.1) can then be written using the relation $\dot{\phi}=Hd\phi/dN$ as
\begin{equation}
H^2=\frac{8\pi V(\phi)}{3m^2(1-4\pi/3m^2(d\phi/dN)^2)}.
\end{equation}
The Hubble parameter is usually considered to be a constant, i.e., $H^2 \cong 8\pi/3m^2V(\phi)$ . In the generalization, however, with $\phi$ being time-dependent, the Hubble parameter is also time-dependent (from equation  (2.3)). Using the variable $N$, equation  (2.2) can be written as
\begin{equation}
\frac{d^2\phi}{dN^2}+\frac{1}{H}\frac{dH}{dN}\frac{d\phi}{dN}+3\frac{d\phi}{dN}+\frac{V^\prime(\phi)}{H^2}=0.
\end{equation}
Equation (2.4) can be further written in terms of $\phi$ and the derivative of $\phi$ with respect to $N$. In order to solve equation  (2.4) numerically, three slow-roll inflation models are adopted: the new inflation model with the potential term given by $V(\phi)=\lambda^2v^4(1-2(\phi/v)^p$ , $(p=3,4)$, the chaotic inflation model with the potential term given by $V(\phi)=M^4/2(\phi/m)^a$ , $(a=2,4,6)$, and the hybrid model $V(\phi)=\alpha[(v^2-\sigma^2)^2+m^2/2\phi^2+g^2\phi^2\sigma^4]\simeq\alpha(v^4+m^2/2\phi^2)$. For example, in the case of chaotic inflation $(a=2)$ using an initial condition of $N=n_0$, $\phi(n_0)=\sqrt{-n_0}/\sqrt{2\pi}m$ and  $d\phi(n_0)/dN=-m/\sqrt{8\pi(-n_0)}$, the numerical solutions to equation  (2.4) for $n_0$ values ranging from -115 to -70 do not reveal any appreciable dependence on $n_0$, and, with the exception of the new inflation model $(p=3)$, this property can be shown for other inflation models. The behavior of $\phi$  for the case in which $n_0 = -100$ is shown in Figure 1 for three inflation models. The inflaton field slowly decreases in the cases of the chaotic inflation model and the hybrid model, but increases very slowly in the case of the new inflation model. The difference of the $N$-dependence (time) of the Hubble parameter among the three inflation models is derived, and the behavior of $H^2(N)/H^2(-60)$  is shown in Figure 2. There is no $N$-dependence in the case of the new inflation model, but the Hubble parameter becomes considerably smaller as $N$ approaches zero in the case of the other models: For example, the value of $H^2(N)/H^2(-60)$ is 1.33 for $N = -80$, 0.835 for $N = -50$, 0.669 for $N = -40$, and 0.504 for $N = -30$, in the case of chaotic inflation $(a=2)$.

The following parameters are used of slow-roll inflation [4]:
\begin{equation}
\epsilon=3\frac{\dot{\phi}^2}{2}(\frac{\dot{\phi}^2}{2}+V)^{-1}=\frac{m^2}{4\pi}(\frac{H^\prime(\phi)}{H(\phi)})^2,
\end{equation}
\begin{equation}
\delta=\frac{m^2}{4\pi}\frac{H^{\prime\prime}(\phi)}{H(\phi)},
\end{equation}
\begin{equation}
\xi=\frac{m^4}{16\pi^2}\frac{H^{\prime}(\phi)H^{\prime\prime\prime}(\phi)}{(H(\phi))^2}.
\end{equation}
Other slow-roll parameters ($\epsilon_V$, $\eta_V$, $\xi_V$) can be written in terms of the slow-roll parameters $\epsilon$, $\delta$, and $\xi$ to the first order in slow roll: $\epsilon=\epsilon_V$, $ \delta=\eta_V-\epsilon_V$, and $\xi=\xi_V-3\epsilon_V\eta_V+3\epsilon^2_V$, where  $\epsilon_V=m^2/16\pi(V^\prime/V)^2$, $\eta_V=m^2/8\pi(V^{\prime\prime}/V)$, and $\xi_V=m^4/64\pi^2(V^\prime V^{\prime\prime\prime}/V^2)$. The behavior of slow-roll parameters $\epsilon$, $\delta$, and $\xi$ with respect to the $N$-dependence differs among the inflation models.
 
The time dependence of the spectral index $n_s$ can also be calculated using these slow-roll parameters. The spectral index is given by 
\begin{equation}
n_s=1+2\delta-4\epsilon.
\end{equation}
The time dependence of $n_s$ is shown in Figure 3 for three inflation models. Figure 3 shows that, for each model, the decreasing behavior of the spectral indexes as well as their values and shapes are different. Taking the new inflation $(p=3)$ as an example, the value of $n_s$ decreases as $N$ approaches zero:  $n_s(-80)=0.950$,$n_s(-60)=0.935$,$n_s(-50)=0.921$, and $n_s(-40)=0.906$ .

\section{Scalar perturbations}

The dependence of the curvature perturbation on the parameter $k$ is investigated assuming a spatially flat Friedman-Robertson-Walker (FRW) universe with a background spectrum given by metric perturbations. The line element for the background and perturbations is generally expressed as [5] 
\begin{equation}
ds^2=a^2(\tau)\{ (1+2A)d\tau^2 -2\partial_i Bdx^i d\tau-[(1-2\Psi)\delta_{ij}+2\partial_i \partial_j E+h_{ij}] dx^i dx^j \},
\end{equation}
where $\tau$ is the conformal time, the functions $A$, $B$, $\Psi$, and $E$ represent the scalar perturbations, and $h_{ij}$ represents tensor perturbations. The density perturbations in terms of the intrinsic curvature perturbation of comoving hypersurfaces is given by $\mathcal{R}=-\Psi-(H/\dot{\phi})\delta\phi$, where $\phi$ is the inflaton field, $d\phi$ is the fluctuation of the inflaton field, and $\mathcal{R}$ is the curvature perturbation. Overdots represent derivatives with respect to time $t$, and primes represent derivatives with respect to the conformal time $\tau$. Introducing the gauge-invariant potential $u\equiv a(\tau)(\delta\phi+(\dot{\phi}/H)\Psi)$ allows the action for scalar perturbations to be written as [6]
\begin{equation}
S=\frac{1}{2}\int d\tau d^3x\{ (\frac{\partial u}{\partial \tau})^2-(\nabla u)^2+\frac{Z^{\prime\prime}}{Z}u^2 \},
\end{equation}
where $Z=a\dot{\phi}/H$ and $u=-Z\mathcal{R}$. 
The field $u(\tau, x)$ is expressed using annihilation and creation operators as 
\begin{equation}
u(\tau , \mathbf{x})=\frac{1}{(2\pi)^{3/2}}\int d^3k\{ u_k(\tau)a_\mathbf{k}+u_k^\ast(\tau)a_{-\mathbf{k}}^\dagger \}e^{-i\mathbf{kx}},
\end{equation}
and the field equation for $u_k(\tau)$ is derived as
\begin{equation}
\frac{d^2u_k}{d\tau^2}+(k^2-\frac{1}{Z}\frac{d^2Z}{d\tau^2})u_k=0,
\end{equation}
where the solution to $u_k$ satisfies the normalization condition $u_kdu_k^\ast/d\tau-u_k^\ast du_k/d\tau=i$. 
Using the slow-roll parameters, $(d^2Z/d\tau^2)/Z$ is written exactly as
\begin{equation}
\frac{1}{Z}\frac{d^2Z}{d\tau^2}=2a^2H^2\nu,
\end{equation}
where
\begin{equation}
\nu=(1+\epsilon-\frac{3}{2}\delta+\epsilon^2-2\epsilon\delta+\frac{\delta^2}{2}+\frac{\xi}{2}).
\end{equation}
A partial integration allows $\tau$ to be expanded as follows [7].
\begin{equation}
\tau=-\frac{1}{aH(1-\epsilon)}-\frac{2\epsilon(\epsilon-\delta)}{aH(1-\epsilon)}+\int\frac{\epsilon(-2\delta^2+\delta(9-4\epsilon)\epsilon-6\epsilon^2+3\epsilon^3-\xi(1-\epsilon))}{(1-\epsilon)^2}\frac{dN}{aH}.
\end{equation}
Here, using the approximate relation $a^2H^2=(1+2\epsilon(\epsilon-\delta))^2/\tau^2(1-\epsilon)^2$ , equation  (3.4) can be rewritten as 
\begin{equation}
\frac{d^2u_k}{d\tau^2}+(k^2-\frac{2\nu(1+2\epsilon(\epsilon-\delta))^2}{\tau^2(1-\epsilon)^2})u_k=0.
\end{equation}
If it is assumed that the slow-roll parameters can be constant, the solution for equation  (3.8) can be written using the Hankel function $H^{(1)}_\mu(-k\tau)$ as
\begin{equation}
u_k(\tau)=\frac{\sqrt{\pi}}{2\sqrt{k}}e^{i(\mu+1/2)\pi/2}(-k\tau)^{1/2}H^{(1)}_\mu(-k\tau),
\end{equation}
where $\mu=\sqrt{2\nu(1+2\epsilon(\epsilon-\delta))^2/(1-\epsilon)^2+1/4}$. This solution is fixed such that as $k\tau\to -\infty $, $u_k(\tau)$ approach plane waves. The numerical solution to equation  (3.4) can then be considered. Equation (3.4) can be written in terms of $N$ as 
\begin{equation}
\frac{d^2u_k}{dN^2}+\frac{1}{aH}\frac{daH}{dN}\frac{du_k}{dN}+(\frac{k^2}{(aH)^2}-2\nu)u_k=0.
\end{equation}
In order to calculate equation  (3.10) numerically, the present-day size perturbation $k=0.002$(1/Mpc) is assumed to exceed the Hubble radius in inflation at the time of $N=-60$. The scale factor is thus written as $a=0.002/H_0e^{N+60}$, where $H_0$ is the value of $H$ at $N=-60$. Since the Hubble parameter and $\nu$ have both been calculated as functions of $N$, equation  (3.10) can be calculated numerically with an initial value of $u_k$ derived from equation  (3.9). Note that the desired solution of $u_k$ is the solution at the point $aH=k$. Taking new inflation $(p=3)$ as an example, in Figure 4, the absolute value of $u_k$ at $aH=k$  derived by numerical calculation of equation  (3.10) is shown as a function of $k$. The absolute value of $u_k$ (equation  (3.9)) obtained using the Bessel approximation, adopting the value of $\mu$ as the numerically calculated value, is also shown. The value of $\mu$ changes slightly according to $N$: $\mu(-80)=1.525$, $\mu(-60)=1.532$, $\mu(-50)=1.540$, $\mu(-40)=1.550$ and $\mu(-30)=1.564$. The results shown in Figure 4 do not suggest any appreciable difference between the numerical calculation and the Bessel approximations, indicating that the Bessel approximation, in which $\mu$ exhibits $k$ dependence, can be considered satisfactory.

The power spectrum of the scalar perturbations $P_\mathcal{R}$ is defined as follows [4].
\begin{equation}
<\mathcal{R}_\mathbf{k}(\tau), \mathcal{R}_\mathbf{l}^\ast(\tau)> = \frac{2\pi^2}{k^3}P_\mathcal{R}\delta^3(\mathbf{k}-\mathbf{l}),
\end{equation}
where, $\mathcal{R}_k(\tau)$ is the Fourier series of the curvature perturbation $\mathcal{R}$. The power spectrum $P_\mathcal{R}$ is then written as [4]
\begin{equation}
P_\mathcal{R} = (\frac{k^3}{2\pi^2})\bigl|\frac{u_k}{Z}\bigr|^2,
\end{equation}
where $Z=a\dot{\phi}/H$. The power spectrum is estimated by three different methods: the numerical method, the Bessel approximation using equation  (3.9), and the Taylor expansion with running (as a familiar method). Calculation of the power spectrum is estimated at $k=aH$. In the case of the Bessel approximation using the values of $\mu$, $d\phi/dN$, and $H$ derived from the numerical calculation (having $k$ dependence), the power spectrum (equation  (3.12)) can thus be written as
\begin{equation}
P_\mathcal{R} = \frac{k^2}{8\pi}\frac{(-k\tau)(H^{(1)}_\mu(-k\tau))^2}{a^2(d\phi/dN)^2}|_{k=aH},
\end{equation}
where the value of $P_\mathcal{R}$ is obtained at $k=aH$ (i.e. $k\tau=-(1+2\epsilon(\epsilon-\eta))/(1-\epsilon))$). There is very little difference between the results of the numerical calculation and the Bessel approximation (see Table 1), again demonstrating that the Bessel approximation is satisfactory. 

The power spectrum is commonly written as 
\begin{equation}
P_\mathcal{R} = \nu^{2\mu-1}(\frac{\Gamma(\mu)}{\Gamma(3/2)})^2(\mu-1/2)^{1-2\mu}\frac{H^4}{m^4|H^\prime|^2}|_{k=aH}.
\end{equation}
The value of the power spectrum $P_\mathcal{R}$  given by equation  (3.14) is $4.0395\times10^6\lambda^2$, assuming the value of $\mu$ at $N=-60$. Using the same parameter values, the numerical calculation by equation  (3.12) and the Bessel approximation (equation  (3.13)) both afford a value $8.5488\times10^6\lambda^2$ . The above values are for the case of the new inflation $(p=3)$, but other inflation models exhibit the same behavior. There is thus a difference of approximately a factor of two between the present numerical and Bessel calculations and the familiar case (equation  (3.14)). This discrepancy can be attributed to the difference in the Hankel function used to derive the expressions. The familiar expressions are derived from the asymptotic form ($k/aH\to 0$) of the Hankel function and are estimated at $4k/aH=1$ [4], whereas the present Bessel calculation uses the correct form. The difference between the asymptotic and correct forms of the Hankel function equates to a factor of approximately $\sqrt{2}$ at $k/aH=1$, giving rise to a change in the normalization of the power spectrum. 

Since the value of the spectral index differs from unity, as indicated by experimental results such as WMAP data [1,8], the power spectrum can be inferred to have $k$ dependence. The power spectrum is usually expanded at a pivot scale $k_0$, i.e. [1],
\begin{equation}
P_\mathcal{R}(k) = P_\mathcal{R}(k_0)(\frac{k}{k_0})^{n_s(k_0)-1+(1/2)\alpha \ln(k/k_0)}
\end{equation}
where $\alpha=dn_s/d\ln k|_{k=k_0}$. Since the three inflation models have non-zero values of $\alpha$, the running case must be considered. If this running term is ignored, with the exception of hybrid inflation, differences with respect to the power spectrum and the angular power spectrum occur between the cases with and without running. Taking the new inflation $(p=3)$ as an example, Table 1 lists the results for the three calculations for various values of $k$, and Figure 5 compares the numerically calculated power spectrum with that determined by the Taylor expansion with power spectrum running, where the overall factor $P_\mathcal{R}(k_0)$ is fixed at $8.5488\times10^6\lambda^2$. These results seem to indicate that the power spectrum given by the Taylor expansion exhibits behavior that differs from that obtained by the present numerical calculation in the case of large $k$. Here, the differences in $k$ dependence of the power spectrum between the Taylor expansion and the proposed numerical method among the inflation models are as follows. A large difference, 1.3 \% $(p=3)$ and 0.8 \% $(p=4)$ occurs at $k = 1$ (1/Mpc) in the case of new inflation. In the case of chaotic inflation, the difference is 0.04 \% in the $\phi^2$ case, 0.3 \% in the $\phi^4$ case, and 0.9 \% in the $\phi^6$ case. The difference in the case of the hybrid model is slight.

Finally, the angular power spectra are calculated using a modified CMBFAST code [9], and likelihood analysis is performed using the WMAP three-year data and the WMAP Likelihood Code [10]. Here, the value of $\chi^2$  is calculated for these inflation models using the same values of the cosmological and inflationary parameters, but only the difference of the value of $\chi^2$ is considered. For the new inflation model, the difference of the value of $\chi^2$ between the familiar Taylor expansion method and the proposed numerical method is nearly 10 for the case in which $p=3$ and is approximately 3.8 for the case in which $p=4$. For the chaotic inflation model, the difference of the value of $\chi^2$ between the familiar Taylor expansion method and the proposed numerical method is approximately zero at $\phi^2$, approximately 1.4 at $\phi^4$ , and approximately 2.7 at $\phi^6$ . However, for the hybrid inflation model, the difference of the value of $\chi^2$ does not occur. The dependence of the initial value $n_0$  in the numerical calculation in the range from -115 to -70 reveals no clear difference, except for new inflation $(p=3)$, where the difference of $\chi^2$ is approximately 2.

\section{Discussion and summary}
We investigate the difference in $k$ dependence of the power spectrum of the curvature perturbation between the familiar Taylor expansion method and a numerical method, and time dependence of the cosmological and inflationary parameters using three slow-roll inflation models. Differences in behavior are observed among the three inflationary models, i.e., the Hubble parameter changes only slightly with time ($N$) for the new inflation but decreases with time for the chaotic inflation model and the hybrid inflation model. The spectral index shows a similar decreasing behavior, but the value and shape are different. The calculation of the power spectrum using the Bessel approximation was shown to be consistent with the numerical calculations, and an approximately two-fold difference in normalization was shown to exist between the present methods and the usual methods (see equation (3.14)) due to the use of the asymptotic form of the Hankel function in the usual treatments. In the $k$-dependent spectra, differences in the power spectrum and the angular power spectrum between the Taylor expansion method and the proposed numerical method occur among the inflation models. Specifically, in the new inflation, a large difference of $\Delta\chi^2\cong10$ occurs. In contrast, in chaotic inflation ($\phi^6$ and $\phi^4$ ), some difference of $\chi^2$ occurs, and in hybrid inflation and chaotic inflation ($\phi^2$ ), the difference of $\chi^2$ rarely occurs. As in the case of the new inflation model ($p=3$), a number of unique and characteristic behaviors, as compared to other inflation models, are derived:
\begin{enumerate}
\item The Hubble parameter changes only slightly in the range from $N=-110$ to $N=-20$.
\item In the numerical calculation, there is a small $n_0$ dependence of the power spectrum and angular power spectrum, but in other inflation models rarely occurs.  
\item The value of $\epsilon$  (see equation (2.5)) is very small, and the value of $|\eta|$ is larger than the value of $\epsilon$.
\item The value of $\xi$, which is not very small, is larger than  $\epsilon$ and oscillates. Moreover, dependence of the initial condition $n_0$  exists.
\end{enumerate}
Unfortunately, the WMAP three-year data was used in the calculation of $\chi^2$. Since the difference of the value of $\chi^2$ is nearly 10, and the values of $\chi^2$ are larger than those for other inflation models in the case of the cosmological and inflationary parameters used herein, it is very interesting what a difference of the best cosmological and inflationary parameters between the Taylor expansion method and the proposed numerical method occurs using the WMAP five-year data.  Moreover, we considered only the case of the scalar perturbation. In the future, we intend to investigate the effect of the tensor perturbation.

\acknowledgments
The authors would like to thank the staff of Osaka Electro-Communication University for their valuable discussions, and Mr. Kouta Asajima for making available a program to calculate $\chi^2$.

\begin{table}[h]
  \caption{Power spectrum values ($P_\mathcal{R} \times 10^6 \lambda^2$) and ratios of these values to those obtained through numerical calculation}
  \label{aaaa}
  \begin{center}
    \begin{tabular}{|c|c|c|c|} \hline
$k$(1/Mpc) & Numerical & Bessel	& Taylor \\ \hline
0.001 &	8.9425	& 8.9413 & 8.9507 \\ \hline
(ratio)	& (1)	& (0.99987) &	(1.0009) \\ \hline
0.002 &  8.5488 & 8.5488 & 8.5488\\ \hline
0.01 & 7.7030 &	7.7011	& 7.6680 \\ \hline
(ratio) & (1) & (0.99974) & (0.99544)\\ \hline
0.1 & 6.5924 & 6.5907 & 6.5305\\ \hline
(ratio) & (1) & (0.99974) & (0.99061)\\ \hline
1.0 & 5.6065 &	5.6052	& 5.5291\\ \hline
(ratio) & (1) & (0.99976) & (0.98619)\\ \hline
    \end{tabular}
  \end{center}
\end{table}

\includegraphics[width=15cm, clip]{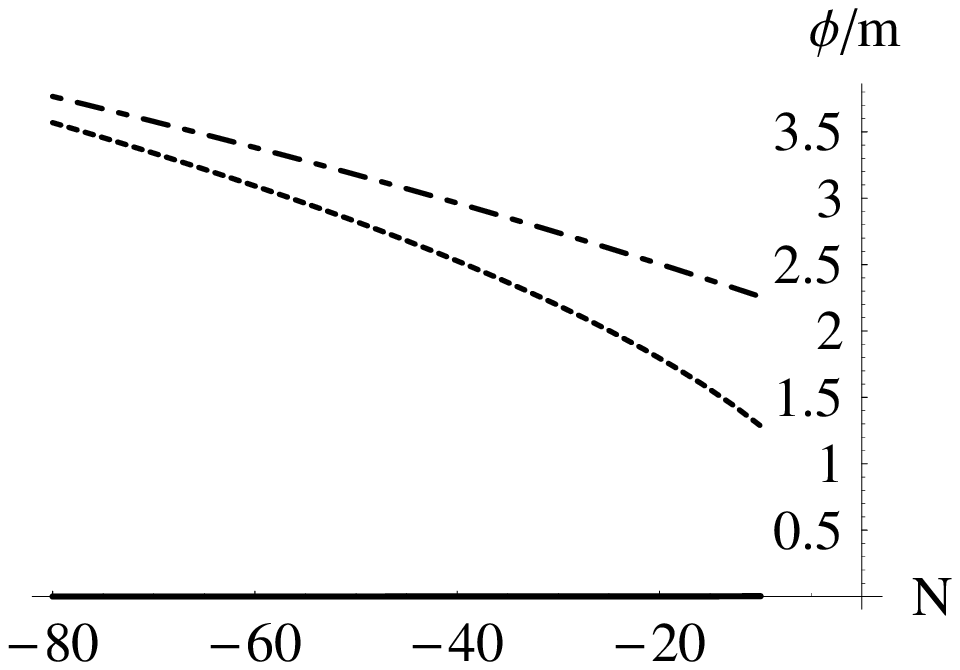}\\
Figure 1: Variation in $\phi/m$ ($m$ is Planck mass) with respect to $N$. The solid line represents new inflation $(p=3)$, the dashed line represents chaotic inflation $(a=2)$, and the dash-dotted line represents hybrid inflation.

\includegraphics[width=15cm, clip]{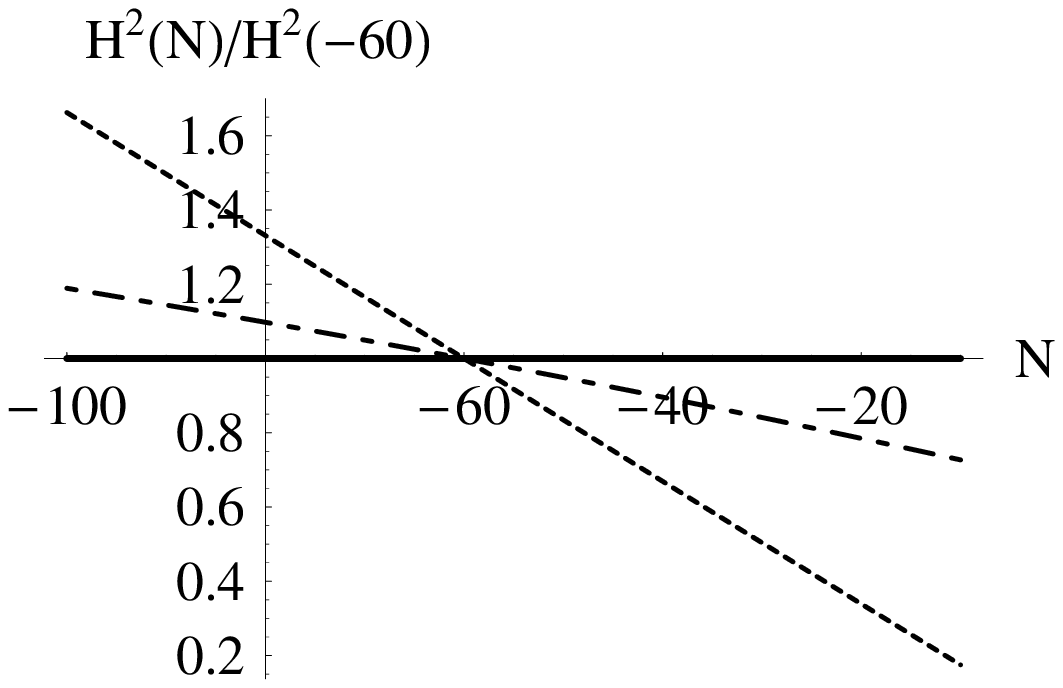}\\
Figure 2: Variation in the Hubble parameter ($H^2(N)/H^2(-60)$) with respect to $N$. The solid line represents new inflation $(p=3)$, the dashed line represents chaotic inflation $(a=2)$, and the dash-dotted line represents hybrid inflation.

\includegraphics[width=15cm, clip]{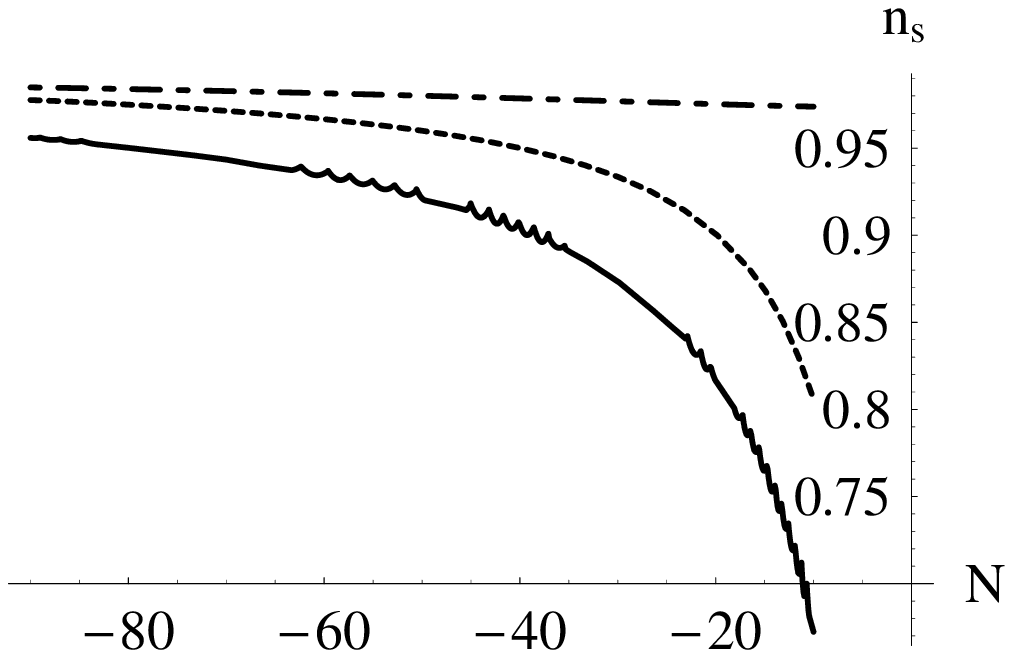}\\
Figure 3: Variation in the spectral index $n_s$ with respect to $N$. The solid line represents new inflation $(p=3)$, the dashed line represents chaotic inflation $(a=2)$, and the dash-dotted line represents hybrid inflation.

\includegraphics[width=15cm, clip]{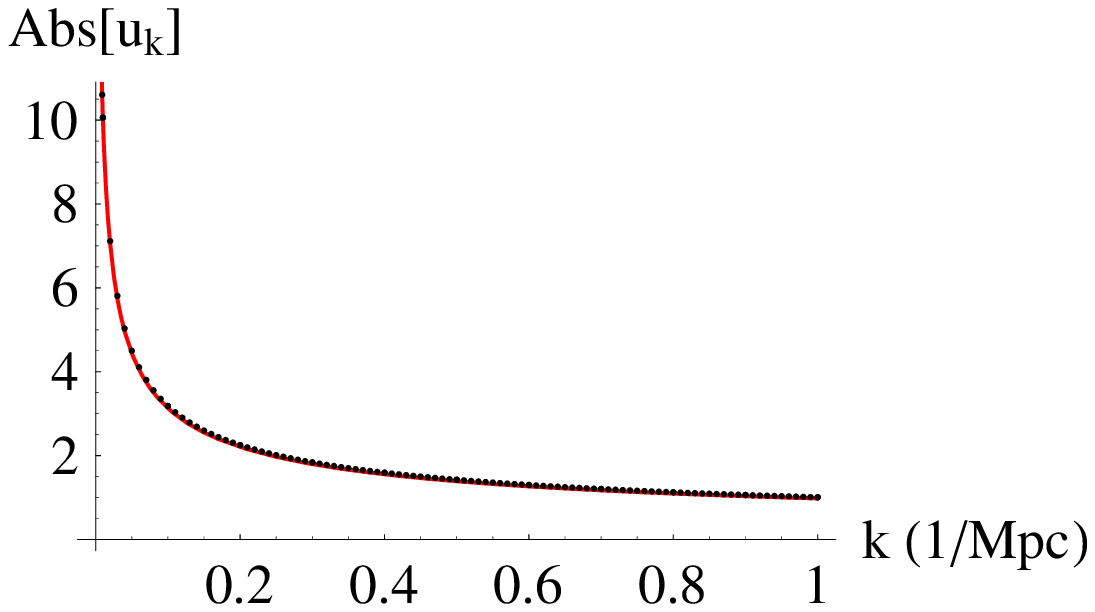}\\
Figure 4: Variation in $u_k$ with respect to $k$. The dotted line represents the numerical calculation, and solid line represents the Bessel approximation using the numerically calculated value of $\mu$ in the new inflation.

\includegraphics[width=15cm, clip]{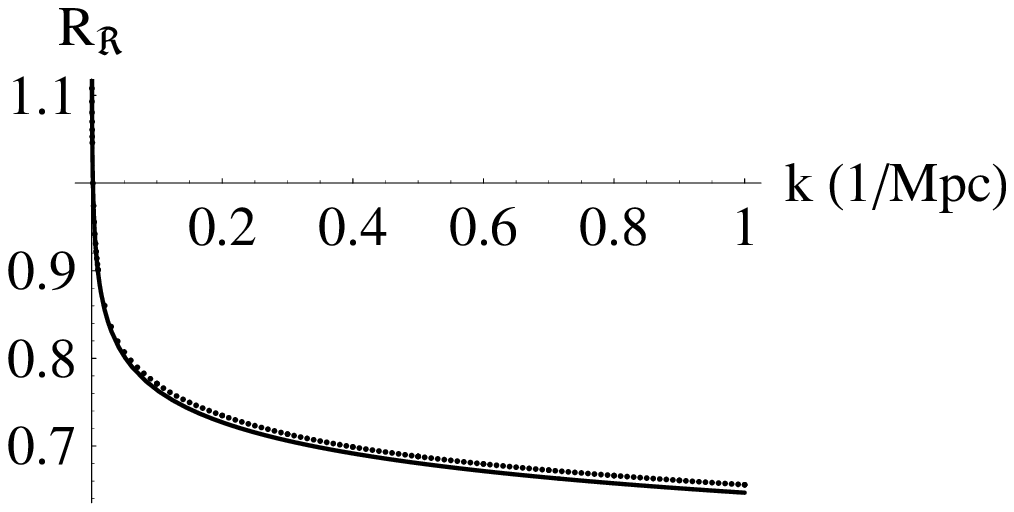}\\
Figure 5: Variation in $P_\mathcal{R}$ with respect to $k$. The dotted line represents the numerical calculation, and the solid line represents the Taylor approximation with spectral index running for the same value of $P_\mathcal{R}(k_0)$ in the new inflation.

\end{document}